# The Quest to Understand the Pioneer Anomaly

Michael Martin Nieto
Theoretical Division (MS-B285)
Los Alamos National Laboratory
Los Alamos, New Mexico 87545 U.S.A

email:   mmn@lanl.gov

**Summary:**  The Pioneer 10/11 missions, launched in 1972 and 1973, and their navigation are reviewed.   Beginning in about 1980 an unmodeled force of ~ 8 x $10^{-8}$ cm/s$^2$ appeared in the tracking data, it later being verified.  The cause remains unknown, although radiant heat remains a likely origin. A set of efforts to find the solution are underway:  a) analyzing in detail all available data, b)  using data from the New Horizons mission, and c) considering an ESA dedicated mission.

## A. The Pioneer Missions

During the 1960's, when the Jet Propulsion Laboratory (JPL) first started thinking about what eventually became the "Grand Tours" of the outer planets (the Voyager missions of the 1970's and 1980's), the use of planetary flybys for gravity assists of spacecraft became of great interest. The idea was to use flybys of the major planets to both modify the direction of the spacecraft and also to add to its heliocentric velocity in a manner that was unfeasible using only chemical fuels. The first time these ideas were put into practice in deep space was with the Pioneers.

Pioneer 10 was launched on 2 March 1972 local time, aboard an Atlas/Centaur/TE364-4 launch vehicle (see Figure 1). It was the first craft launched into deep space and was the first to reach an outer giant planet, Jupiter, on 4 Dec. 1973 [1,2]. Later it was the first to leave the "solar system" (past the orbit of Pluto or, should we now say, Neptune). The Pioneer project, eventually extending over decades, was managed at NASA/AMES Research Center under the hands of four successive project managers, the legendary Charlie Hall, Richard Fimmel, Fred Wirth, and the current Larry Lasher.

While in its Earth-Jupiter cruise, Pioneer 10 was still bound to the solar system. By 9 January 1973 Pioneer 10 was at a distance of 3.40 AU (Astronomical Units[1]), beyond the asteroid belt. This in itself was a happy surprise, as the craft had not been destroyed passing through. With the Jupiter flyby, Pioneer 10 reached escape velocity from the solar system. It was then headed in the general direction opposite the relative motion of the solar system in the local interstellar dust cloud or opposite to the direction towards the galactic center.

Pioneer 11 followed soon after with a launch on 6 April 1973, cruising to Jupiter on an approximate heliocentric ellipse. This time during the Earth-Jupiter cruise, it was determined that a carefully executed flyby of Jupiter could put the craft on a trajectory to encounter Saturn in 1979. On 2 Dec. 1974 Pioneer 11 reached Jupiter, where it underwent the Jupiter gravity assist that sent it back inside the solar system to catch up with Saturn on the far side. It was then still on an ellipse, but a more energetic one. Pioneer 11 reached as close to the Sun as 3.73 AU on 2 February 1976.

Pioneer 11 reached Saturn on 1 Sept. 1979. The trajectory took the craft under the ring plane on approach and it came within 24,000 km of Saturn. After encounter, Pioneer 11 was on an escape hyperbolic orbit. The motion of Pioneer 11 is approximately in the direction of the Sun's relative motion in the local interstellar dust cloud (towards the heliopause). Its direction is roughly anti-parallel to the direction of Pioneer 10.

In Figure 2 the trajectories of the Pioneers in the inner solar system are shown. In Figure 3 the trajectories of the Pioneers ands Voyagers over the entire solar system are shown.

## B. The Pioneer Navigation

The navigation to Jupiter was carried out at the Jet Propulsion Laboratory using NASA's Deep Space Network (DSN). It was ground-breaking in its advances and fraught with crises. To succeed the navigation team needed to modify the codes with real-time fixes. (See Figure 4.) But they succeeded.

---

[1] An Astronomical Unit is the mean Sun-Earth distance, about 150,000,000 km.

The navigation used a Doppler signal. An S-band signal (~2.11 Ghz) was sent via a DSN antenna located either at Goldstone, California, outside Madrid, Spain, or outside Canberra, Australia. On reaching the craft the signal was transponded back with a (240/221) frequency ratio (~2.29 Ghz), and received back at the same station (or at another station if, during the radio round trip, the original station had rotated out of view). There the signal was de-transponded by (221/240) and any Doppler frequency shift was measured directly by cycle count compared to an atomic clock. The idea was to determine the velocity as a function of time and from this calculate a trajectory, a procedure that is done iteratively to improve the accuracy.

However, to obtain the spacecraft velocity as a function of time from this Doppler shift is not easy. The codes must include all gravitational and time effects of general relativity to order $(v/c)^2$ and some effects to order $(v/c)^4$. The ephemeredes of the Sun, planets and their large moons as well as the lower mass multipole moments are included. The positions of the receiving stations and the effects of the tides on the exact positions, the ionosphere, troposphere, and the solar plasma are included.

Given the above tools, precise navigation was possible because, due to a serendipitous stroke of luck, the Pioneers were spin-stabilized. This is contrary to, for example, the later Voyagers which were 3-axis stabilized. With spin-stabilization the craft are rotated at a rate of ~(4-7) rpm about the principal moment-of-inertia axis. Thus, the craft is a gyroscope and attitude maneuvers are needed only when the motions of the Earth and the craft move the Earth from the antenna's line-of-sight. With 3-axis stabilization, there are continuous, semi-autonomous, small gas jet thrusts to maintain the antenna facing the Earth. This yields a navigation that is not as precise as that of the Pioneers.

The Pioneers were the first deep spacecraft to use nuclear heat from $^{238}$Pu as a power source in Radioisotope Thermoelectric Generators (RTGs). The RTGs were placed at the end of long booms to be away from the craft and thereby avoid any radiation damage. (See Figure 5.) Thus, the craft had to be spin-stabilized. Especially in the later years, only a few orientation maneuvers were needed every year to keep the antenna pointed towards the Earth, and these could be easily modeled.

Even so, there remained one relatively large effect on this scale that had to be modeled: the solar radiation pressure of the Sun, which also depends on the craft's orientation with respect to the Sun. This effect is approximately 1/30,000 that of the Sun's gravity on the Pioneers and also decreases as the inverse-square of the distance. It produced an acceleration of ~20 x $10^{-8}$ cm/s$^2$ on the Pioneer craft at the distance of Saturn (9.38 AU from the Sun at encounter). (For comparison, the gravitational acceleration of the Sun at the Earth is 0.593 cm/s$^2$.) Therefore, any "unmodeled force" on the craft could not be seen very well below this level at Jupiter. However, beyond Jupiter it became possible.

**C. Discovery of the Anomaly**

One of the main experiments on the Pioneers was radioscience celestial mechanics. In 1969 John Anderson became the PI of this program, remaining so until the official end of the extended mission [3] in 1997. Working with Eunice Lau (who also later joined the Pioneer anomaly Collaboration), the Pioneer Doppler data going back to 1976 for Pioneer 11 and 1981 for Pioneer 10 (but also including the Jupiter flyby) was archived at the National Space Science Data Center (NSSDC), something that later was extremely helpful.

Part of the celestial mechanics effort, working together with the navigation team, was to model the trajectory of the spacecraft very precisely and determine if there were any unmodeled effects. Around Jupiter none could be found. But over time, a number of approximately 6-month to 1-year averages of the data were taken from both Pioneer 10 and Pioneer 11 and by 1987 it was clear that an anomalous acceleration appeared to be acting on the craft with a magnitude ~ 8 x $10^{-8}$ cm/s$^2$, directed approximately towards the Sun. (See Figure 6.)

For independent reasons, in 1994 the current author contacted Anderson about gravity in the solar system. When the anomaly came up its magnitude was a great personal surprise. The result was an announcement in a 1994 Conference Proceedings. The strongest immediate reaction was that the anomaly could well be an artifact of JPL's Orbital Data Program (ODP), and could not be taken seriously until an independent code had tested it. So Anderson put together a team that included two former Pioneer co-workers (see Figure 4) who were then associated with The Aerospace Corporation. These two used the independent CHASMP navigation code they had developed to look at the Pioneer data. To within small uncertainties, their result was the same.

The Pioneer anomaly Collaboration's discovery paper appeared in 1998 [4] and a final detailed analysis appeared in 2002 [5]. The latter used Pioneer 10 data spanning 3 January 1987 to 22 July 1998 (when the craft was 40 AU to 70.5 AU from the Sun) and Pioneer 11 data spanning 5 January 1987 to 1 October 1990 (when Pioneer 11 was 22.4 to 31.7 AU from the Sun). The result, after accounting for all known systematics, was that there is an unmodeled acceleration, directed approximately towards the Sun, of

$$a_P = (8.74 \pm 1.33) \times 10^{-8} \text{ cm/s}^2.$$

**D. Meaning of the anomaly**

The decision to use modern data in the final analysis was motivated by a number of reasons. i) It was easily accessible and in modern format, ii) the craft were then further away from the Sun (greater than 40 and 20 AU, respectively, for Pioneers 10 and 11) so solar radiation pressure was a smaller complicating factor, and iii) further out there were fewer antenna Earth-reorientation manoeuvres that had to be modeled. To the accuracy of the analysis, the anomaly was constant, but this accuracy was only ~15%.

This brings up the problem of heat radiating out from the craft in a non-isotropic manner. Since at launch there were 2500 W of heat coming from the RTGs and only 63 W of directed power could cause the effect, it is tempting to assume this must be the cause. However, even though admittedly this is the most likely explanation of the anomaly, no one as yet has been able to firmly tie this down, despite heated controversy [6]. The craft was designed, again serendipitously, so that the heat was radiated out in a very fore/aft symmetric manner. Further, the heat from electric power went down by almost a factor of 3 during the mission. Heat as a mechanism remains to be clearly resolved.

Drag from normal matter dust as well as gravity from the Kuiper belt have been ruled out. Also, if this is a modification of gravity, it is not universal; i.e., it does not affect planetary bodies in bound orbits. It could, in principle be i) some strange modification of gravity, ii) drag from dark matter or a modification of inertia, or iii) a light acceleration. (Remember, the signal is a Doppler shift which is only *interpreted* as an acceleration.) In such circumstances the true direction of the anomaly should be i) towards the Sun, ii) along the craft velocity vector,[2] or iii) towards the Earth. (If the origin is heat the acceleration would be iv) along the spin axis.)

**E. Finding the origin of the anomaly**
*a) Using all the data.* If all the Doppler data, from launch to last contact, were to be analyzed together a number of things would be obviated [7]. First, it would be easier to see if the anomaly is truly a constant or rather if it exhibits a half-life corresponding to the 87.74 years of $^{238}$Pu. Also, if the effects of solar radiation pressure and many manoeuvres that occur close in to the Sun could be disentangled, one might be able to do 3-dimensional tracking precisely enough to determine the exact direction of the anomaly. Perhaps most intriguingly, by closely studying the data around Pioneer 11's Saturn flyby (and Pioneer 10's Jupiter flyby) it could be determined if, indeed, there was an onset near these transitions to hyperbolic-orbits.

The Doppler data archived at the NSSDC and the data used in the summary analysis [5], as well as other pieces obtained elsewhere have recently been reacquired, translated and compiled in modern format. Analyses will soon start [8]. The Doppler data holds the possible key to finding an origin to the anomaly.

Due to the foresight of Larry Kellogg of Ames in retaining obsolete telemetry files, the engineering data has also been reacquired [8]. In the long run the telemetry might be most useful. From the beginning the Collaboration has observed that, even if the anomaly turns out to be due to systematics, the anomaly inquiry would still result in a win. One would obtain a better understanding of how to build spacecraft for very deep space and how to model and track craft there.

*b) The New Horizons mission to Pluto*
On 19 Jan 2006 the New Horizons mission to Pluto and the Kuiper Belt was launched from Cape Canaveral. Although it was not designed for precision tracking, it might be able to yield useful information.

The first problem will be the on-board heat systematics. The large RTG is mounted on the side of the craft, and produced ~4,500 W of heat at launch. A rough calculation shows that a systematic of ~20 cm/s$^2$ or larger will be produced. Since the post-launch modeling of heat systematics is notoriously difficult, this makes this systematic an important problem to overcome.

---
[2] Technically it is along the vector sum of the spacecraft velocity and the dark matter's change in velocity.

A saving grace may be that soon after launch a 180 degree "Earth acquisition manoeuvre" rotation was performed, to aim the main antenna at the Earth. The difference in the Doppler shift immediately before and after the rotation can in principle yield a difference measurement of the heat acceleration which would be pointed first in one direction and then in the opposite. But a determination may be difficult because of the high solar radiation pressure (which will vary somewhat in the two orientations) and the relatively small data set before the manoeuvre.

More gratifyingly, New Horizons will be in spin-stabilization mode for about the six months before the Jupiter observing period (January-June, 2007, with encounter on 28 Feb. 2007). It also will be spin-stabilized for much of the period after June 2007 until soon before the Pluto encounter on 14 July 2015. This is designed to save fuel so it can be used to aim later at a Kuiper Belt Object. With luck the Doppler and range data from these periods will supply a test, at some level, of the Pioneer anomaly, especially since the velocity of the craft before (~21 km/s) and after (~25 km/s) the Jupiter encounter will be significantly different that those of the Pioneers (~12 km/s). Perhaps something can be learned from the New Horizons data by 2008.

*c) ESA's Cosmic Vision*
As discussion on the anomaly was proceeding, in Europe there independently arose an international interest in the problem. In May 2004 a meeting was held at the University of Bremen to discuss the anomaly, and from this an international Pioneer Explorer Collaboration was formed to propose a dedicated test of the anomaly [9], with Hansjoerg Dittus of Bremen as PI. Institutions from all over Europe, including from France, Germany, Great Britain, Italy, Netherlands, Portugal, and Spain, have joined.

The proposal is a Theme for ESA's Cosmic Vision program, with launches to occur during the period 2015-2025. As such it's timing would be perfect if the two investigations described above indicate that a dedicated test of the anomaly is called for. A driving consideration would be new technology. The mission would take insight from knowledge of what allowed the Pioneer craft to be navigated so well and add to it.

The concept would be to determine accurately the heliocentric motion of a test-mass utilizing 2-step tracking with common-mode noise rejection.[3] A state-of-the-art Ka-band tracking system, using both Doppler and range, could be used to track the main satellite to an accuracy approaching $0.1 \times 10^{-8}$ cm/s$^2$. Then, from the forward side facing away from the Earth, a "formation flying" system would send out small corner-cube covered spheres to be tracked from the main satellite with mW laser ranging. The passive spheres would be at a distance of order >500 m from the main satellite, which satellite would utilize occasional manoeuvres to maintain formation. This final step could yield an acceleration precision approaching $10^{-10}$ cm/s$^2$. On board one could also carry sensitive drag-free DC accelerometers, which are being developed.

The craft would be spin-stabilized. The design would be extremely fore/aft symmetric as far as heat/power-radiation were concerned, to reduce heat acceleration of the craft. (The heat would be radiated out in fore/aft and axially symmetric manners.) In Figure 7 we show a schematic cut-away preliminary model. The side exterior surfaces are curved to symmetrically reflect and radiate heat from the side of the bus. (Heat from inside the equipment bus would come out of louvers located between the RTG extensions). This also symmetrically reflects heat from the RTGs which are in parabolic Winston cone reflectors. The RTGs could be extended out on booms after launch. One can also see where the spheres would be extruded and the central location of the laser.

---

[3] Another concept would be an autonomous probe that would be jettisoned from a main vehicle, such as the InterStellar Probe. This would happen further out than at least the orbit of Jupiter or Saturn. The probe would then be navigated from the ground.

The test masses would not be released until the main craft would undergo no further acceleration manoeuvres, be it from a final stage chemical rocket, a planetary flyby, or even a jettisoned solar sail. This would probably be at a distance of 5-10 AU, when the craft hopefully had a velocity of >5 AU/yr. From then on, and especially at distances of 25-45 AU, when solar radiation pressure is reduced, precise data could be taken.

**F: Conclusion**
That the Pioneer anomaly is a physical effect is no longer in doubt. The only question is its origin. Here the anomaly's discovery and the growing interest and efforts to understand it have been described The latter include, in order of possible completion, a) an analysis of (almost) the entire Pioneer Doppler data set, b) the possibly fruitful analysis of the tracking data from the ongoing New Horizons mission to Pluto, and c) a dedicated ESA mission. Understanding the anomaly will yield, at the least, improved navigational protocols for deep space and, at the best, exciting new physics.

Finally, given this opportunity, I wish to more directly address the current audience with a political question that Europe must face. Up to now Europe has not ventured into deep space alone. Its greatest triumph, Huygens, necessitated a piggy-back on a RTG-powered NASA mission, Cassini. Why is this? Because of the political mine-field about anything nuclear.

There is simply no way, given any foreseeable near-term technology, that even a medium-sized spacecraft (few hundred kg) can go into deep space (>5 AU) in a short time (less than a few years) without some form of on-board nuclear power (the simplest being RTGs).

However, it is my experience that the elder statesmen of Europe are very hesitant to even discuss the matter. (Recall that they reached maturity during the anti-nuke era.) On the brighter side, I have found that young post-docs have much more of an "of course" attitude towards using RTGs. At the least, I hope that the Pioneer Explorer proposal will help stimulate discussion on the matter. If there is not a shift in the European paradigm, then Europe will end up abandoning deep space to the rest of the world. That would be extremely sad.

**FIGURE CAPTIONS:**

**Figure 1:** Pioneer 10's launch on 2 March 1972.

**Figure 2:** The Pioneer orbits in the interior of the solar system.

**Figure 3:** Ecliptic pole view of the Pioneer 10, Pioneer 11 and Voyager trajectories. Pioneer 10 is traveling in a direction almost opposite o the galactic center, while Pioneer 11 is heading approximately in the closest direction to the heliopause. The direction of the solar system's motion in the galaxy is approximately towards the top.

**Figure 4:** Members of the JPL navigation team working on the day of Pioneer 10's launch. In the foreground are Tony Liu and Phil Laing, who later became part of the Pioneer anomaly Collaboration. In the background are Sun Kuen Wong, Jack Hohikian, Steve Reinbold, and Bruce O'Reilly. Note the stack of computer program cards labeled "LAST CARD," the large format computer printout paper, and the Tektronix scope, evidence of the technologies used then.

**Figure 5** A diagram of the Pioneer spacecraft. The final fins on the RTGs were actually larger, to increase the heat radiation away from the craft.

**Figure 6:** A JPL Orbital Data Program (ODP) plot of the early unmodeled accelerations of Pioneer 10 and Pioneer 11, from about 1981 to 1989 and 1977 to 1989, respectively. This graph first appeared in JPL memos from the period 1992.

**Figure 7:** A schematic, cut-away drawing of the Pioneer Anomaly Explorer concept. The side facing away contains the radio antenna to communicate with Earth. On the facing side are the canisters that will emit corner-cube covered spheres and the mW laser. (Drawing courtesy of Alexandre D. Szames.)

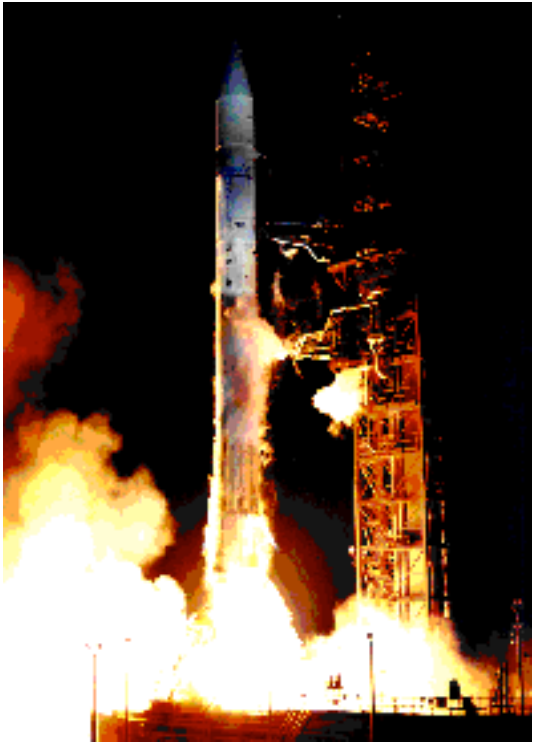
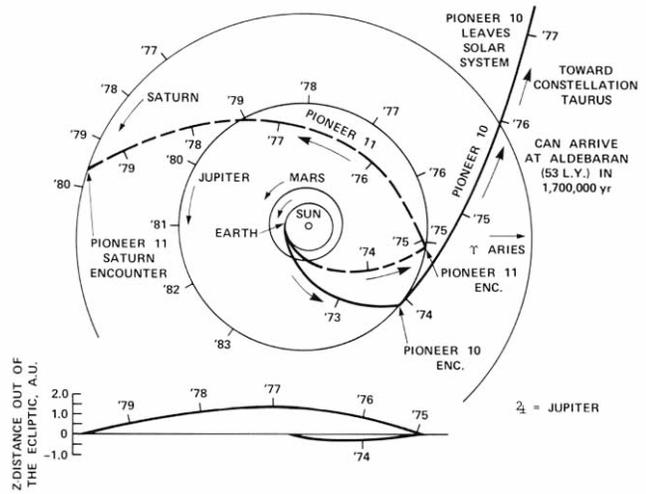
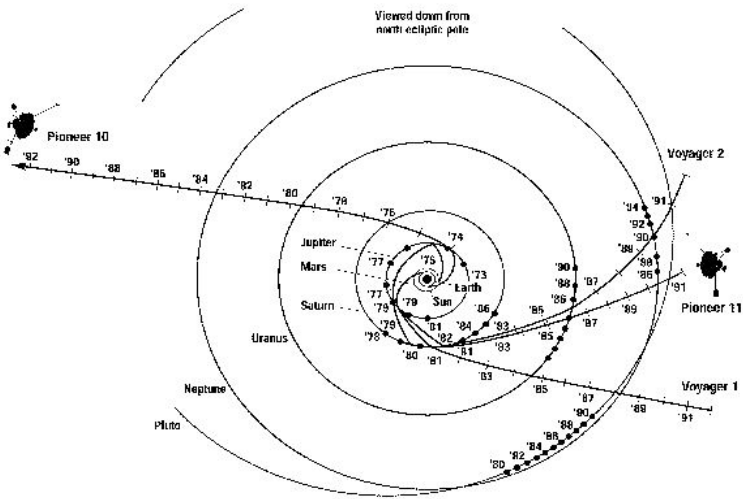

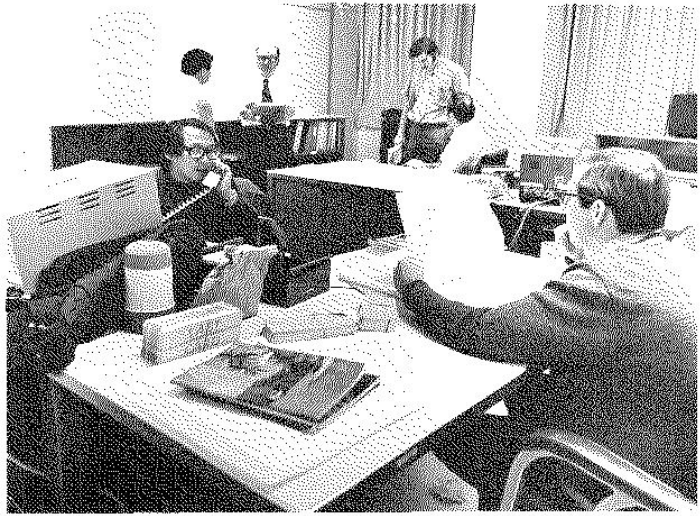

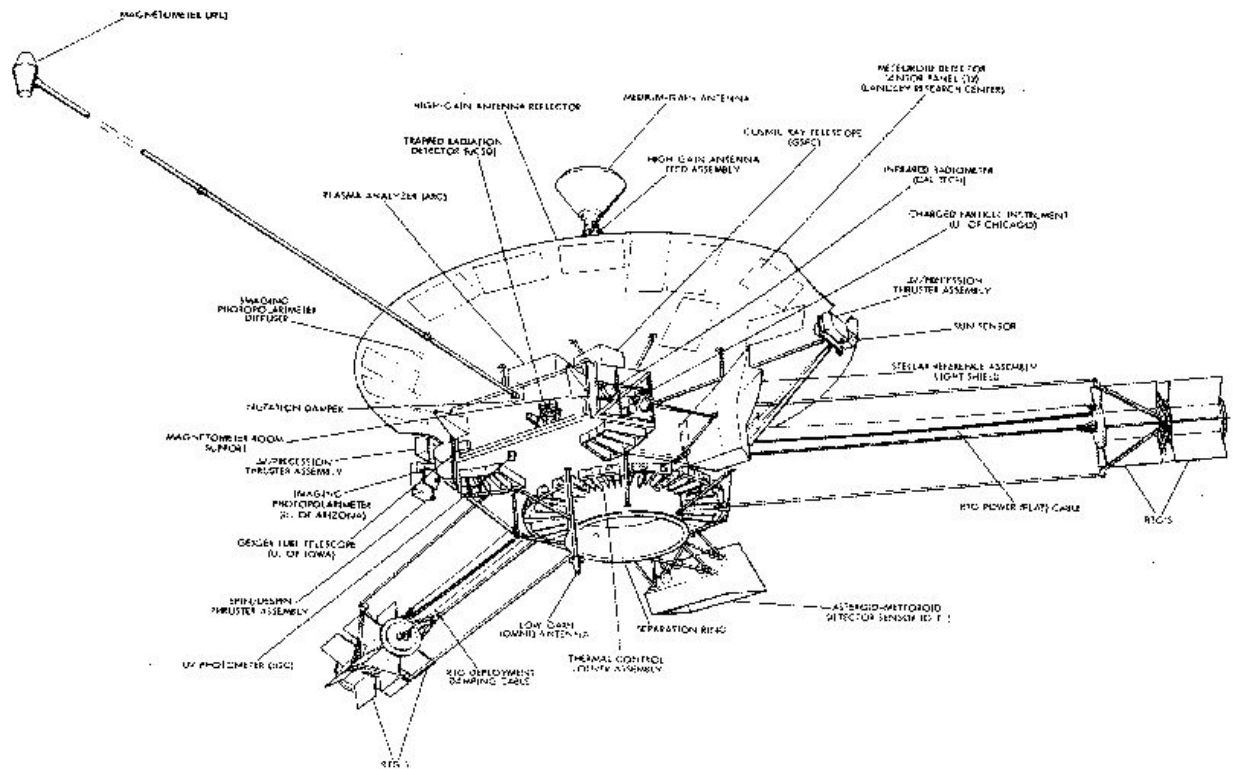

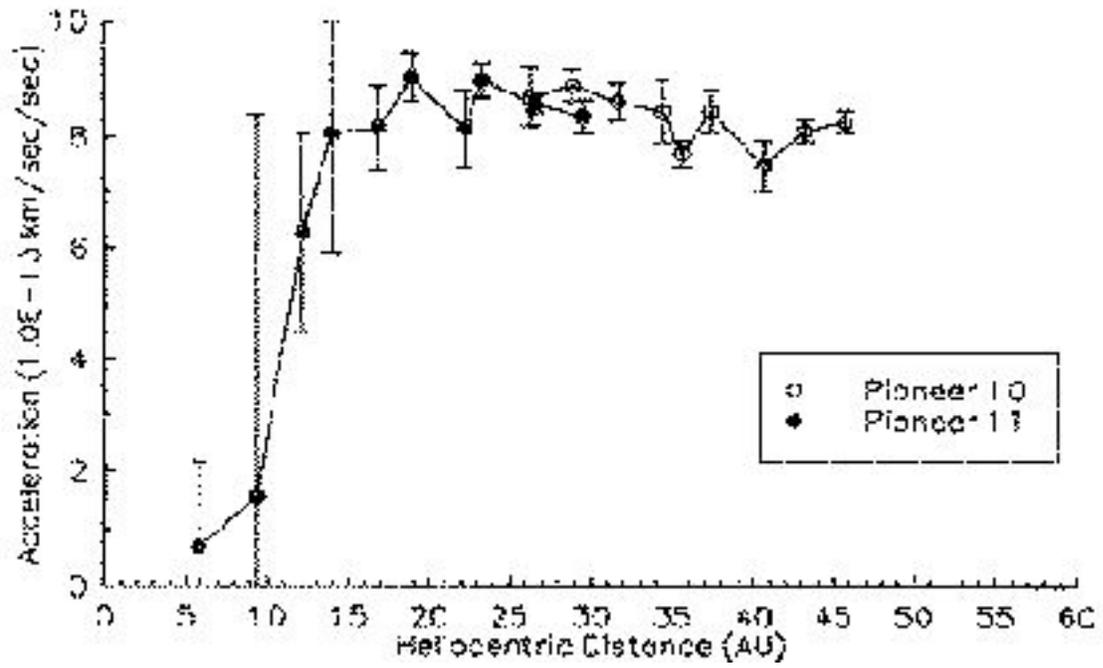
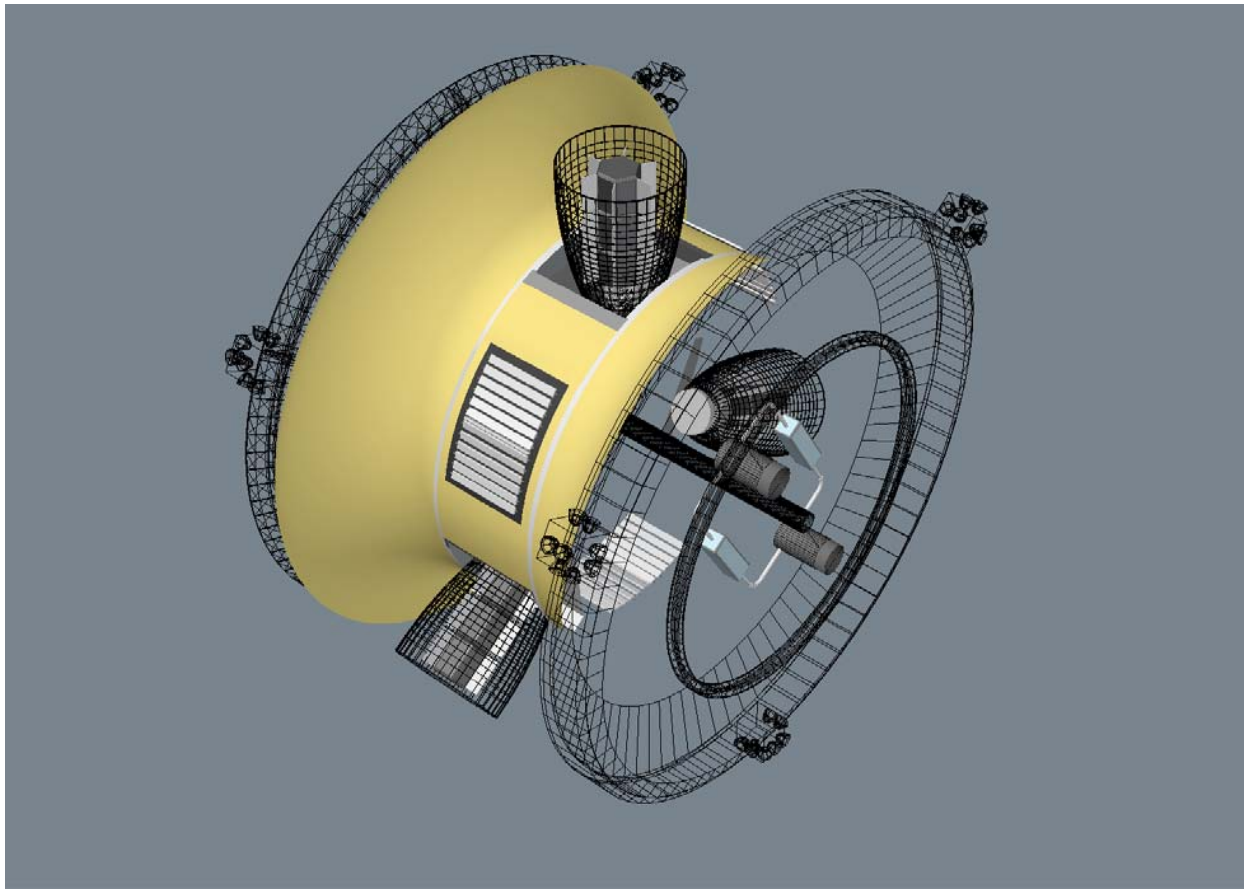